**Title:** Continuous production for large quantity plasma activated water using multiple plasma device setup

**Authors**

Vikas Rathore[1,2*], Chirayu Patil[1], and Sudhir Kumar Nema[1,2]

1. Atmospheric Plasma Division, Institute for Plasma Research (IPR), Gandhinagar, Gujarat 382428, India

2. Homi Bhabha National Institute, Training School Complex, Anushaktinagar, Mumbai 400094, India

**\*Email: vikas.rathore@ipr.res.in**

**Abstract**

In the present work, a batch and continuous production of plasma-activated water (PAW) is reported. To produce PAW in a batch and continuous manner a multiple plasma device setup is used. The multiple plasma device consists of a series of plasma devices that are powered simultaneously to produce PAW. This multiple plasma device is powered by indigenously developed high-voltage high-frequency power supply. The air plasma generated in this multiple plasma device setup is electrically characterized and the produced radicals/species are identified using optical emission spectroscopy. The post-discharge effluent gases left after plasma-water exposure carries some environmental pollutants ($NO_x$ and $O_3$, etc.). The batch and continuous PAW production setup utilizes effluent (pollutants) gases in production of large volume PAW. Hence, it substantially reduces the concentration of these pollutants in effluent gases which are released in environment.

The batch process produces high reactive PAW with less volume (2 liters). Moreover, in a continuous process, a high volume (20 liters) with low reactivity of PAW is produced. The high reactive PAW and low reactive PAW are used for different applications. Inactivation of microbes (bacteria, fungi, viruses, and pests), food preservation, selective killing of cells, etc. is carried out using high reactive PAW whereas low reactive PAW has applications in seeds germination, plant growth, and as a nitrogen source for agriculture and aquaculture applications, etc. In addition, the batch and continuous PAW production setup designs are scalable, therefore, can be used in industries for PAW production.

## 1. Introduction

The plasma activated water (PAW) technology is a step toward sustainable development. It is an eco-friendly and economically viable technology that utilizes plasma and water as preliminary ingredients. PAW has applications in microbial inactivation (bacteria, fungi, viruses, algae, and pests, etc.)(1-6), food preservation (fruits, vegetables, dairy products, and meat products including seafood, etc.)(7-12), dental and medical equipment's surface disinfection(13), pesticide removal(14), selective killing of cancer cells(15), virus vaccine preparation(16), seeds germination(17), plant growth(18), plasma acid(19), and as a source of nitrogen fertilizer(20), etc.(21, 22) Hence use for disinfectants (chemicals), preservatives, fertilizer, and medicine ingredients, etc. that cause various environmental and health issues can be avoided. Since, they can cause soil and water pollution that causes various diseases in plants and animals (land and water) including humans(23-26). Hence, PAW has the potential to be used as a replacement of disinfectants (chemicals), preservatives, fertilizer, and medicine ingredients, etc. Therefore, PAW technology has the potential to be used in sustainable development.

The mentioned applications of PAW are possible because of the presence of numerous reactive oxygen-nitrogen species (RONS) in PAW(1, 2, 4, 7, 13, 15-18, 22). Some of the RONS present in PAW are given as $NO_3^-$ ions, $NO_2^-$ ions, $H_2O_2$, OH radical, $ONOO^-$ ions, and dissolved $O_3$, etc.(4, 7, 17, 19, 27-31) The presence of reactive oxygen species (ROS) such as $H_2O_2$, OH radical, $ONOO^-$ ions, and dissolved $O_3$, etc. plays a significant role in applications like microbial inactivation, food preservation, and selective killing of cancer cells, etc.(1, 2, 4, 6, 7) Moreover, a higher concentration of reactive nitrogen species (RNS) like $NO_3^-$ ions and $NO_2^-$ ions, etc. exists in the form of nitric and nitrous acids. As a result, pH of PAW decreased significantly which also favors applications like microbial inactivation and food preservation, etc.(1, 2, 7, 9)

Moreover, low to moderate concentration of RONS present in PAW is used in applications like seed germination and plant growth(12, 17, 32). The PAW treatment improves the wettability properties of seeds by removing the waxing from their surface. Moreover, low to moderate ROS concentration acts as a signalling molecule that enhances plant growth(17, 32). In addition, the RNS present in PAW can also be used as a nitrogen source for applications in agriculture and aquaculture(22). Hence, the concentration of RONS in PAW decides its applications. The present work also emphasizes the production of PAW with low/moderate and high RONS concentrations.

The agriculture and aquaculture applications of PAW required a large quantity of PAW. Moreover, applications like microbial inactivation and food preservation, etc. required a relatively low volume of PAW. Also, limited research has been conducted which emphasizes the production of PAW in large quantities. Jin et al.(30) reported mass production of PAW using a cylindrical dielectric barrier discharge reactor with very high plasma discharge power (8 kW). Also, they used a cooling assembly to cool down electrodes which created additional running costs. They showed the variation in pH, electrical conductivity (EC), and $NO_3^-$ ions concentration in PAW only. Moreover, Ĉech et al.(3) use hydrodynamic cavitation plasma jet (HCPJ) for mass production of PAW. However, Ĉech et al.(3) do not report the change in physicochemical properties and RONS concentration in PAW. Hence, commenting on the reactivity of plasma activated water becomes difficult. Past literature of numerous authors mainly reported the production of small volume PAW using cold plasma with low plasma discharge power (1W to 40 W)(5, 10, 13-15, 21, 32). Hence, production of high volume of PAW using low plasma discharge power still need to be investigated.

The insubstantial literature on high volume PAW production using cold plasma with low plasma discharge power and its properties analysis become the motivation of the present work. Therefore, the present work emphasizes the production of high volume of PAW using cold plasma with low plasma discharge power (~ 35 W). Also, tries to optimize the dissolution of generated plasma species/radicals in water by trapping gases reactive species/radicals carried by effluents gases after PAW production(33). The post-discharge effluent gas after plasma-water exposure carries environmental pollutants such as $NO_x$ and gases $O_3$ etc (34-36). Hence, the reduction of these gases' pollutants concentration in effluent gases becomes important. Moreover, trapping of these gases using PAW further enhances the physicochemical properties and RONS concentration in it(33). As a result, maximum utilization of discharge gases species is possible.

The present work discussed scalable, in terms of volume of plasma activated water setup using multiple plasma devices powered by an indigenously developed power supply. These setups include batch production of PAW and continuous production of PAW. In batch production, a comparatively low volume (2 liters) of PAW with high reactivity and RONS concentration tries to be achieved. In continuous production, a high volume (20 liters) of PAW with moderate reactivity and RONS concentration tries to be achieved. The importance of high and low/moderate reactivity PAW in applications is already discussed above. The voltage-current waveform is used to characterize the plasma. The generated air plasma species are diagnosed by using optical emission spectroscopy. The PAW characterization is performed by measuring the physicochemical properties of PAW and RONS concentration in it.

## 2. Material and methods

2.1 Optical emission and electrical diagnosis of multiple plasma device setup

Figure 1 (a) showed the schematic of multiple plasma device setup. A series of five plasma devices were used in multiple plasma device setup. The plasma devices are connected in parallel with the power supply. The plasma device is a cylindrical co-axial assembly in which quartz cone was used as dielectric, mesh as a ground electrode, and copper pipe as power electrode(28). The quartz double cone has a two-sided B24 male socket. A flexible copper wire mesh was wrapped around the cone to make a ground electrode. The power electrode was made using a 16 mm outer diameter copper pipe of with a thickness of 2 mm. A diamond knurling of pitch size of 0.5 mm was introduced which disturbed the uniform electric field and increased the localized electric field at the sharp knurled edges. More detail about plasma device can be found in our past reported work (28). A self-made high voltage high frequency (HVHF) and low current power supply was used to power multiple plasma device setup. The voltage drops across the multiple plasma device setup was measured using a high voltage 1000x probe (Tektronix P6015A) and a 4-channel, 100 MHz, 2 GS s$^{-1}$ sampling rate digital oscilloscope (Tektronix TDS2014C). The total current and transported charge in multiple plasma device setup were measured using a voltage probe (Tektronix TPP0201) and an oscilloscope as shown in figure 1 (a)(37). This probe measured the voltage drop across a 3-ohm non-inductive resistor and 2.2 µF non-polar capacitor connected in series with the ground.

The air plasma emission spectrum in the afterglow region produced in the plasma device was recorded using optical fiber and spectrometer (Model EPP2000-UV from StellarNet Inc.) in the wavelength range of 200-600 nm as shown in figure 1 (a). The captured light from the radiative decay of atoms and molecules was transmitted using optical fiber to the charge-coupled device (CCD) detector of the spectrometer with a detector integration time of 500 ms. The spectrometer has 1200 lines mm$^{-1}$ grating groove density, 60 mm effective focal length, and a 0.5 nm of spectral resolution. The calibration of the spectrometer was performed using an emission line of wavelength 253.6 nm emitted from a mercury vapour lamp.

2.2 Batch process for PAW production

The schematic of batch production of plasma activated water is shown in figure 1 (b). The multiple plasma device setup was loosely fit over the flat bottom flasks (B24 socket fitted). That substantially reduces the discharge gas leakage. Each flat bottom flask (500 capacity) carries 400 ml (5-flasks, total water volume of 2 liters) of ultrapure milli-Q water (control). Air was fed from the top using an air pump at a constant flow rate of 5 l min$^{-1}$ using an air rotameter. The plasma-water exposure time for batch production varies between 0.5 hours to 4 hours. The properties (RONS and physicochemical properties) of PAW was monitored periodically. The discharge gases effluent after plasma-water exposure from flask-I feed to flask-II, so the undissolved gases reactive in flask-I get dissolved in flask-II and so on. The series trapping of reactive species present in effluent gases enhances the physicochemical properties and dissolved RONS concentration in PAW. Furthermore, it reduces the risk of release of environmental pollutants species carried by post-discharge effluent gases such as $NO_x$ and $O_3$ to the open atmosphere. A schematic of the series trapping assembly of gases reactive species in water is shown in figure 1 (b, c).

2.3 Continuous process for PAW production

The schematic of the continuous production of plasma activated water is shown in figure 1 (c). In a continuous process, water was recirculated between the flask during plasma-water exposure. The total volume of water taken in the continuous process was 20 liters kept in a closed water tank. A water pump was used to feed the water from the tank to the flask. The level of water in flasks is controlled using inlet and outlet control values. After the water activation cycle is over, the activated water returns to the water tank and feedback again to the flask as shown in figure 1 (c). Furthermore, periodic monitoring of PAW was performed to study the change in its properties for up to 10 hours. The remaining operating condition was kept constant similar to the batch process.

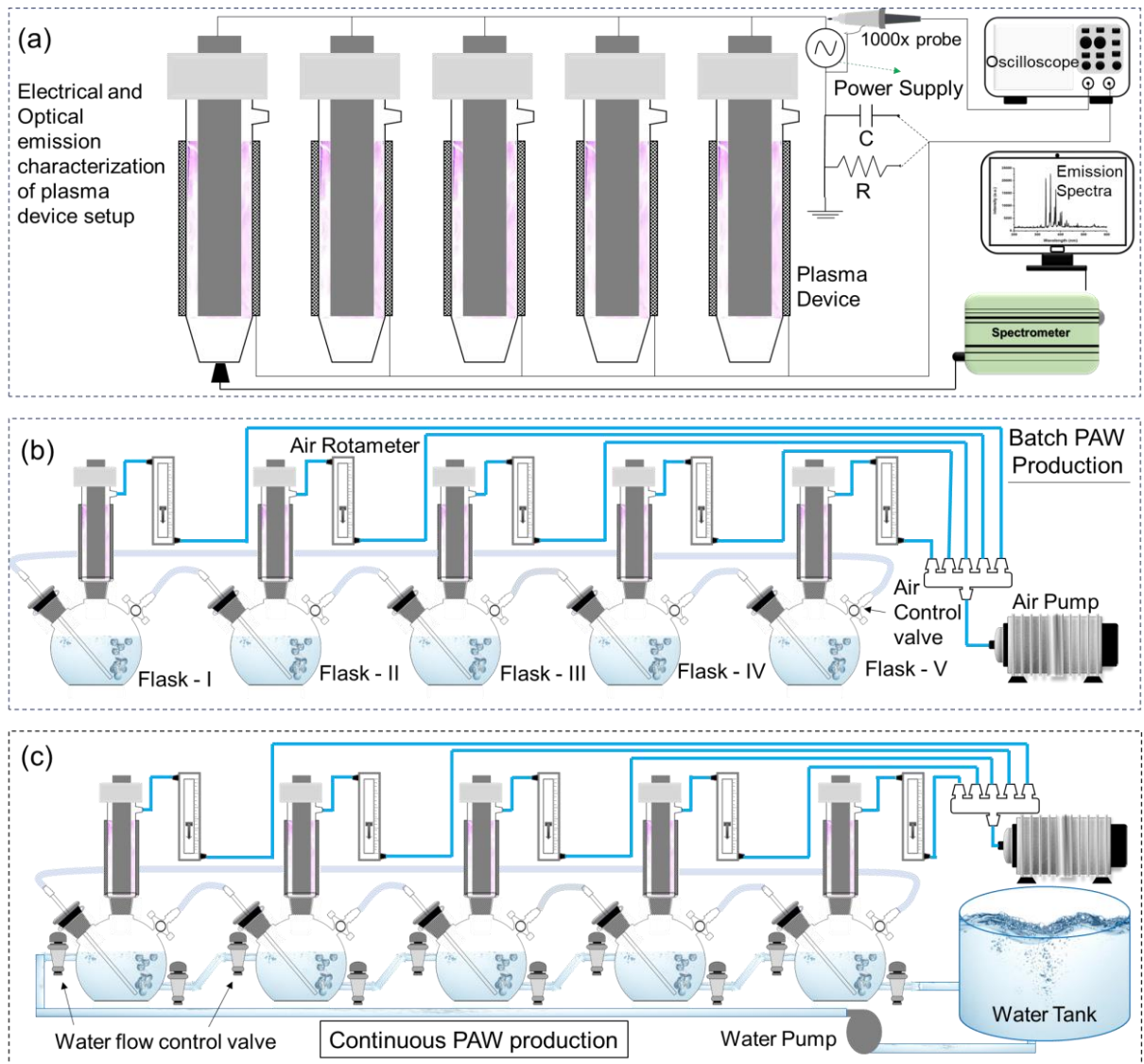

Figure 1. (a) Schematic of electrical and optical emission characterization of air plasma produced using multiple plasma device setup. (b) Schematic of batch production of plasma activated water using multiple plasma device setup. (c) Schematic of continuous production of plasma activated water using multiple plasma devices setup

2.4 Schematic of power supply

Figure 2 showed the schematic of the power supply used in the batch and continuous production of plasma activated water using multiple plasma device setup. The input to the power supply was single-phase (1-ϕ) 50 Hz 220 V alternative current (AC) power. The input AC is converted to direct current (DC) using a full-wave diode bridge rectification circuit. The filter circuit converts the rectified DC to pure DC as shown in Figure 2. The DC voltage was lies between 0 to 300 V. The received DC were converted to high frequency pulsating quasi sine wave output voltage by designing a high-frequency inverter circuit. The inverter circuit was developed using a full-bridge Pulse Width Modulator (PWM) topology with a high switching frequency oscillator (OSC). The high-frequency pulsating quasi sine wave output was connected to the primary side of the high voltage high frequency (HVHF) transformer. The HVHF transformer transforms 0-300 V to 0-10 kV output voltage with tuneable output frequency of 0-30 kHz (38-40). The high voltage (HV) output coming from the secondary side of the transformer followed by a filter was used as input to multiple plasma device setup to produce plasma activated water.

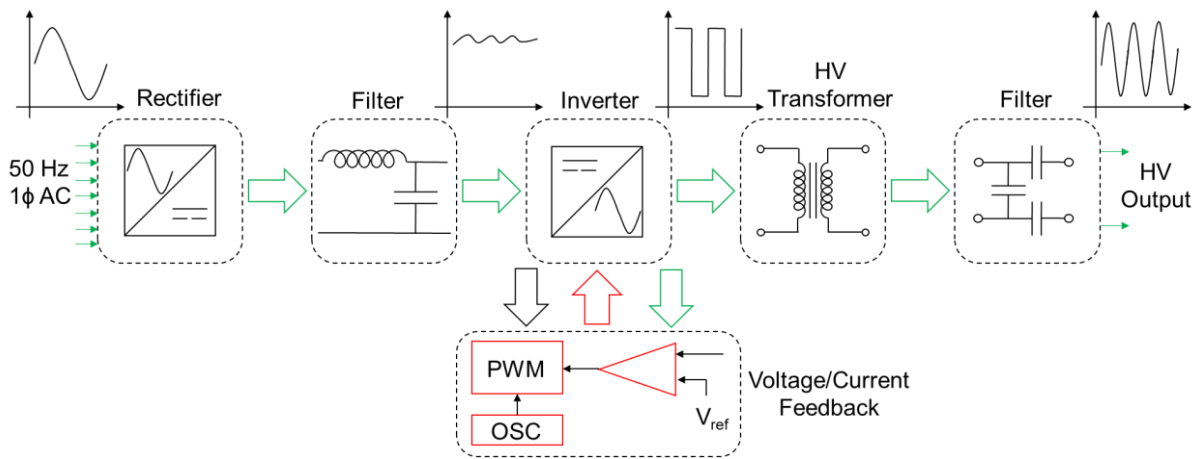

Figure 2. Schematic of power supply used for PAW production using multiple plasma device setup

2.5 Measurement of properties of PAW

The physicochemical properties namely pH, oxidation reduction potential (ORP), total dissolved solids (TDS), and electrical conductivity (EC) of PAW were determined using a Hanna Instruments pH meter (HI98121), Hanna Instruments ORP meter (ORP-200), HM Digital TDS meter (AP1), and HM Digital EC meter (COM-360).

The reactive oxygen nitrogen species (RONS) concentrations namely $NO_3^-$ ions, $NO_2^-$ ions, $H_2O_2$, and dissolved $O_3$ concentration in PAW were initially determined semi-quantitatively. The semi-quantitative estimation of RONS was performed using strips test and colorimetry test kits. The strips used for the estimation of $NO_2^-$ ions and $H_2O_2$ concentration were given as QUANTOFIX Nitrite and QUANTOFIX Peroxide 25 (MACHEREY-NAGEL). The colorimetry test kits used to determine $NO_3^-$ ions and dissolved $O_3$ were given as VISOCOLOR alpha Nitrate (MACHEREY-NAGEL) and Dissolved Oxygen Chemical Test Kit - HI3810 (Hanna Instruments).

The UV-visible spectroscopy (SHIMADZU UV-2600)(37) were used to measure the RONS concentration in PAW. The $NO_3^-$ ions concentration in PAW was determined by measuring the absorbance of PAW at 220 nm (Deuterium lamp) and using a standard calibration curve of $NO_3^-$ ions (molar attenuation coefficient at wavelength 220 nm is 0.0602 $(mg\ L^{-1})^{-1}\ cm^{-1}$)(37). The $NO_2^-$ ions present in PAW (acidic region) when reacts with the reaction mixture sulfanilamide and N-(1-naphthyl)ethylenediamine dihydrochloride give radish purple azo dye and showed peak absorbance at 540 nm (Tungsten lamp). The standard curve of $NO_2^-$ ions (molar attenuation coefficient at wavelength 540 nm is 0.0009 $(\mu g\ L^{-1})^{-1}\ cm^{-1}$) was used to determine the unknown concentration of $NO_2^-$ ions(37). Similarly, $H_2O_2$ is present in PAW (acidic region) when reacts with titanium (IV) ions to form peroxotitanium (pertitanic acid). That formed a yellow color complex which showed maximum absorbance at 407 nm (Tungsten lamp). The unknown $H_2O_2$ concentration in PAW was determined using a standard $H_2O_2$ curve having a molar attenuation coefficient of 0.4857 $mM^{-1}\ cm^{-1}$(37). Moreover, the $NO_2^-$ ion presence in PAW interferes with $H_2O_2$ determination. Since, the $NO_2^-$ ions react with $H_2O_2$ and suppress its concentration beyond the detection limit(27, 37). Therefore, to inhibit the $NO_2^-$ ions and $H_2O_2$ reaction, azide ions ($N_3^-$) were added to PAW as soon as it was prepared. The $N_3^-$ ions react with $NO_2^-$ ions and for $N_2$ gas, as a result, $NO_2^-$ ions interference in $H_2O_2$ concentration determine can be avoided. The indigo colorimetric method was used to determine the dissolved $O_3$ concentration in PAW (37, 41). The expression to calculate the unknown dissolved $O_3$ in PAW is given as:

$$\frac{mg}{l} \ of \ O_3 = \frac{100 \times \Delta A}{f \times b \times v} \tag{1}$$

*ΔA* – difference in absorbance of PAW and control at 600 nm (Tungsten lamp)

*f* – sensitivity factor (0.42)

*v* – volume of sample in ml

b – optical path length of cell (cm)

2.6 Data Analysis

All the experiments shown in present study were repeated at least three times (n ≥ 3). The data collected were expressed as means ± standard deviation (μ ± σ). The statistically significant difference with a p-value (null hypothesis significance test) of 0.05 between the observed results were calculated using one-way analysis of variance (ANOVA) followed by a post-hoc test (Tukey's Honest Significant Difference (HSD)).

**3. Results and Discussions**

3.1 Characterization of air plasma and identification of plasma species

The voltage-current waveform of air plasma is shown in figure 3 (a). The current shown in figure 3 (a) is a combination of continuous alternating current and discontinuous discharge current. The discharge current is the current associated with generated radicals and species in the plasma. These charged particles carry discharge current in each rising and falling current half-cycle. Figure 3 (a) shows the discharge current is in the form of a combination of several filamentary current peaks. Hence, this discharge is known as filamentary micro discharge.

Therefore, the air plasma produced in plasma devices was DBD filamentary micro-discharge in nature(42). The energy and power consumed in multiple plasma device setup were given as 1.74 mJ and 34.8 W calculated using the charge voltage Lissajous figure, respectively.

The emission spectrum of the air plasma afterglow region (the optical fiber capturing light photons placed 10 mm away from plasma discharge region in which mainly neutral molecules and atoms are present) is shown in figure 3 (b). The air plasma showed strong emission band peaks of nitrogen gas ($N_2$) second positive system (SPS). The observed vibrational band peaks in afterglow air plasma showed the transition of $N_2$ (C $^3\Pi_u$) higher state to $N_2$ (B $^3\Pi_g$) lower state. In addition, a weak intensity $N_2^+$ first negative system (FNS) was also observed in air plasma afterglow regions. The observed vibrational band peaks of $N_2^+$ FNS showed the transition from $N_2^+$ (B $^2\Sigma_u^+$) higher state to $N_2^+$ (X $^2\Sigma_g^+$) lower state(43, 44). The radiative decay of $N_2$ and $N_2^+$ higher energy state to lower energy state results in the formation of the mentioned electronic vibrational transition shown in figure 3 (b).

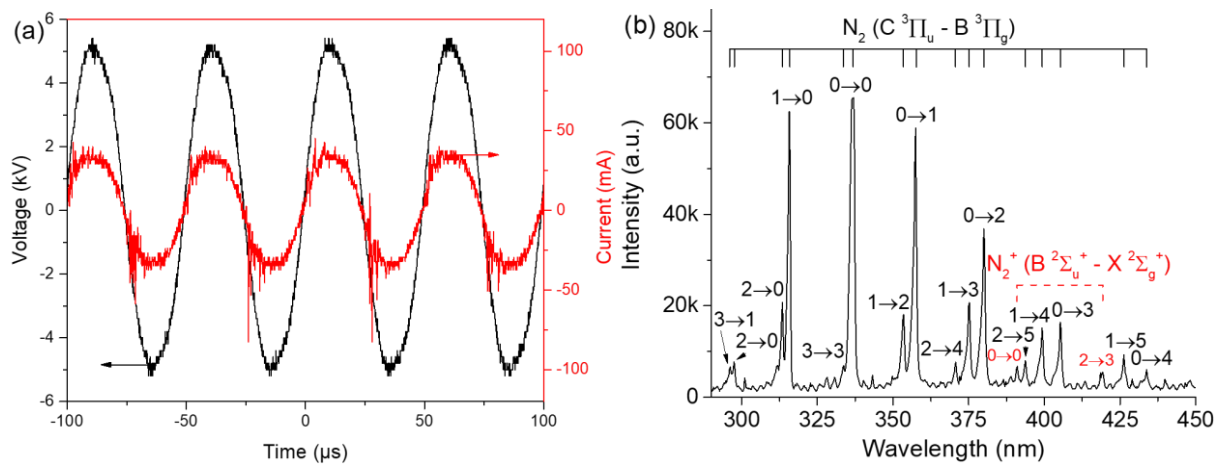

Figure 3. (a) Voltage-current characterization of multiple plasma device setup. (b) Emission spectrum of air plasma produced in plasma device

3.2 Physicochemical properties of PAW prepared in batch process

The charge radicals and species formed due to air discharge in plasma devices when exposed to water formed various stable and unstable reactive radicals and species in water. The high energy identified species observed in the air plasma afterglow region participate in various gases reactions which formed species like N, H, $e^-$, NO, $NO_2$, OH, $H_2O_2$, $HO_2$, $H_2O$, O, $O_2$, and $O_3$, etc. are shown in equations (2-9) of appendix Table A1 (27, 28, 31, 45-48).

These generated species get dissolved in water and form various reactive oxygen-nitrogen species (RONS) in water. The formation of various RONS in PAW is shown in equations (10-21) of appendix Table A1 (27, 28, 31, 45-52). The presence of these RONS in PAW responsible for the physicochemical changes observed in PAW. Figure 4 showed the variation in physicochemical properties of PAW with plasma treatment time kept in the flasks – I to V.

The occurrence of $NO_2^-$ and $NO_3^-$ ions, etc. in PAW exist in form of nitrous and nitric acid due to which the decrease in pH of PAW observed (1, 2, 5, 30, 37). Increasing plasma-water treatment time significant ($p < 0.05$) decreases the pH of PAW as shown in figure 4 (a). The decreasing pH signifies the increasing concentration of acidic species present in PAW with increasing plasma-water exposure time. In addition, no statistically significant ($p > 0.05$) difference was observed in PAW prepared in different flasks. The percentage decrease in pH of PAW of flasks (flask-I to flask-V) compared to control after 4-hours of plasma-water exposure time varied between 55.2% to 62.7%, respectively. The observed trend of decrease in pH of PAW with increasing plasma-water exposure time is also supported by previously reported work of Ten Bosch et al.(5), Punith et al.(21), Sajib et al.(32), Jin et al.(30), Xiang et al.(6), and Subramanian et al.(15), etc.

The dissolution of various oxidizing species ($H_2O_2$, dissolved $O_3$, OH, $ONOO^-$, $NO_2^-$ ions, and $NO_3^-$, etc.) in PAW increases the oxidizing potential (ORP) of PAW (6, 12, 16, 28,

29, 37). The variation in ORP of PAW with increasing plasma-water exposure time is shown in figure 4 (b). Increasing plasma-water exposure time substantially ($p < 0.05$) increased the ORP of PAW kept in flasks I to V. The ORP of PAW kept in the flasks (I to V) varies between 550 mV to 605 mV which was 139.1% and 163.0% higher compared to control after 4-hours of plasma-water exposure. The past reported work of Xiang et al.(6) also confirmed an increase in oxidizing tendency (ORP) of PAW with increasing plasma treatment time.

The generated dissolved radicals/species in PAW increased the total dissolved solids (TDS) and electrical conductivity (EC) of PAW. The inorganic ions such as $NO_2^-$ ions and $NO_3^-$ ions etc. present in PAW makes the PAW electrically conducting. Figure 4 (c, d) reveals increasing plasma treatment time with water increases the TDS and EC of PAW. This signifies a continuous increase in the concentration of dissolved conducting species in PAW with longer plasma-water exposure. For 4-hours of plasma treatment time, the highest value of TDS and EC observed in PAW were given as 600 ppm and 1610 µS cm$^{-1}$, respectively. The increase in EC with increasing plasma-water exposure time was also shown by Xiang et al.(6), Jin et al.(30), Subramanian et al.(15), and, Than et al.(18), respectively.

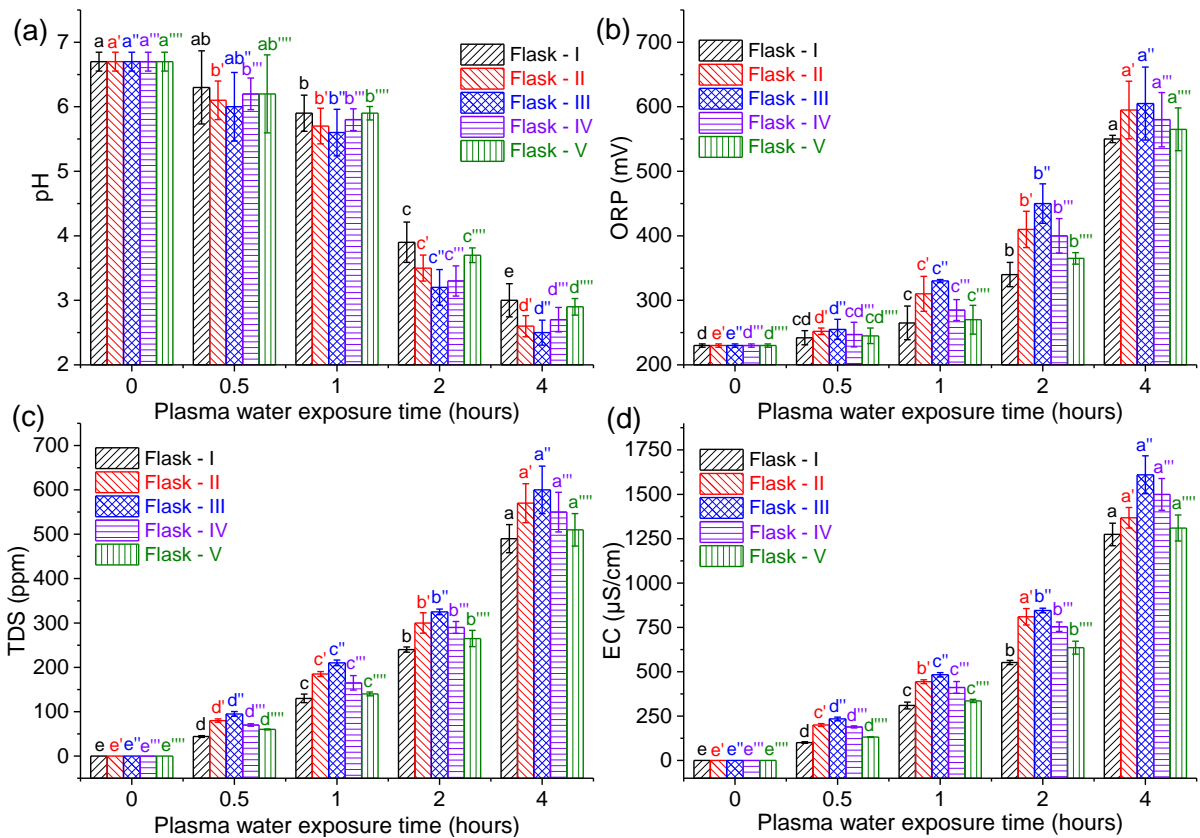

Figure 4. Physicochemical properties of plasma activated water prepared using multiple plasma device batch setup (refer figure 1 (b)). The statistically significant difference (p < 0.05) among the different properties of PAW groups with time (mean ± standard deviation) (μ ± σ) is shown by different lowercase letters.

3.3 RONS concentration in PAW prepared in the batch process

The formation of RONS in PAW is shown in equations (10-21) of appendix Table A1. Some of the stable RONS ($NO_3^-$ ions, $NO_2^-$ ions, $H_2O_2$, and dissolved $O_3$) present in PAW were identified using semi-quantitative estimation and their actual concentrations were measured using UV-visible spectroscopy(37). Figure 5 showed the measured RONS concentration in PAW. Initially, in the control (plasma-water exposure time 0 hours), no dissolved RONS were present in PAW (figure 5). As the plasma exposure to water was initiated, the formation of RONS in water starts occurring. Increasing plasma-water exposure time increases the $NO_3^-$ ions, $NO_2^-$ ions, and $H_2O_2$ concentration in PAW (figure 5 (a-c)). These RONS concentration variations with time are also supported by work reported by Subramanian et al.(15), Xiang et al.(6), Jin et al.(30), Sajib et al.(32), Lukes et al.(27), Punith et al.(21), and, Than et al.(18). However, the concentration of dissolved $O_3$ present in PAW initially increased and then decreased with increasing plasma-water exposure time. This showed the dissolved $O_3$ present in PAW reacts with other RONS (equations (11, 13) of appendix Table A1) that results in suppression of its concentration (figure 5 (d)). Moreover, the $NO_2^-$ ions and $H_2O_2$ also reacted with each other and other RONS present in PAW by following equations (19) of appendix Table A1 (27). However, $NO_2^-$ ions and $H_2O_2$ did not show a decrease in their concentration with time. In addition, the observed maximum concentration of $NO_2^-$ ions and $H_2O_2$ in PAW were 10.7 mg $L^{-1}$ and 6.7 mg $L^{-1}$. This was substantially low compared to the maximum concentration of $NO_3^-$ ions which was 176.0 mg $L^{-1}$. This showed the stability of $NO_3^-$ ions (equations (11, 16, 20, 21) of appendix Table A1) over other RONS in PAW. This was due to other RONS ($NO_2^-$ ions, dissolved $O_3$, $H_2O_2$) reacted with each other to form more stable $NO_3^-$ ions. As a result, the $NO_3^-$ ions concentration in PAW was substantially higher compared to $NO_2^-$ ions, $H_2O_2$, and dissolved $O_3$, respectively. The substantially higher concentration of $NO_3^-$ ions in PAW relative to rest RONS is also shown in past literature by various researchers(5, 6, 15, 28, 29, 32, 33, 37).

In the batch production process, a 4-hour plasma-water exposure gave PAW of ORP up to 605 mV and pH up to 2.5. The high ORP and low pH of PAW showed antimicrobial efficacy of PAW. Since the high oxidizing PAW has the efficacy to oxidize the pathogenic microbes and low pH supports this antimicrobial efficacy of PAW(6). Hence, the prepared PAW could be used in applications like food (fruits, vegetables, daily products, and meat products including seafood) preservations, surface disinfection including surgical equipment, dentistry, and pathogens (bacteria, fungi, viruses, and pests, etc.) inactivation, etc.(1-13)

In conclusion, the batch process could be used to prepare PAW of high reactivity that could be used in numerous applications. A prototype (scale-up) of the present batch model could be used in industries to prepare PAW of high reactivity with a larger volume.

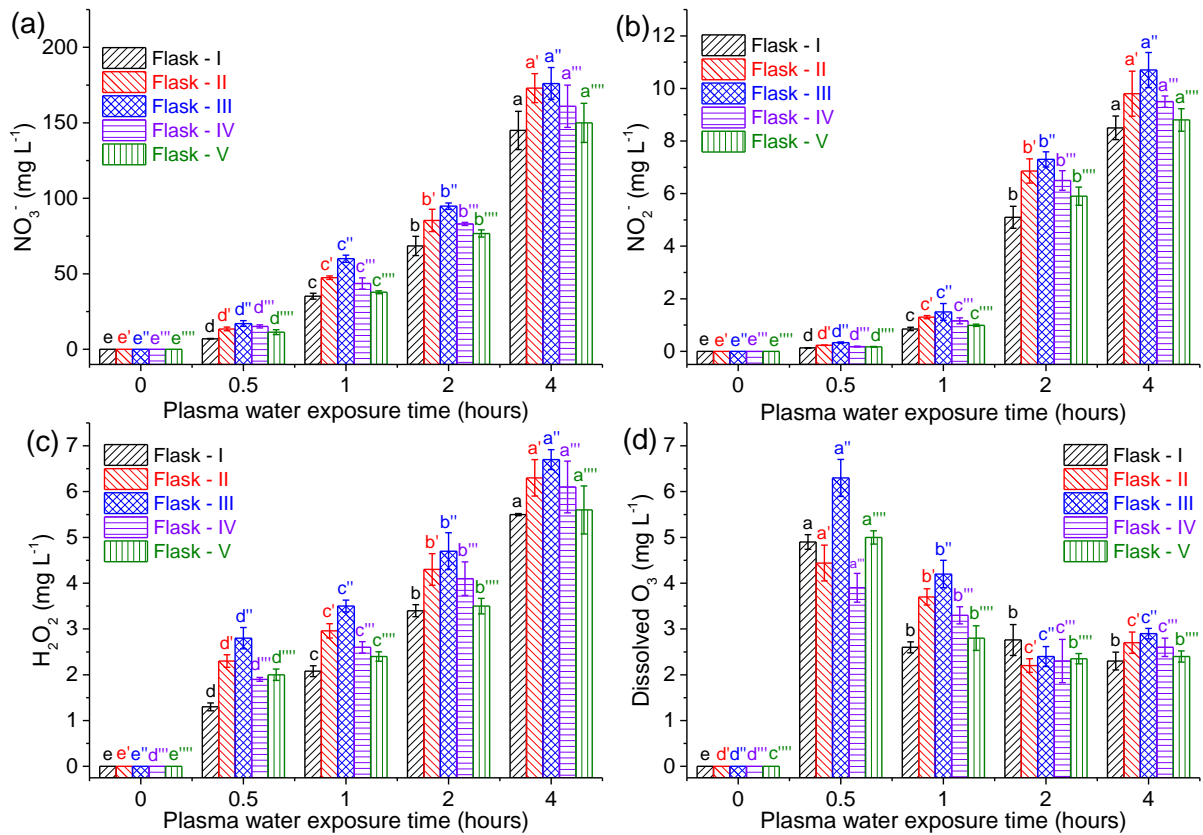

Figure 5. Reactive oxygen-nitrogen species concentration ((a) $NO_3^-$ ions, (b) $NO_2^-$ ions, (c) $H_2O_2$, and (d) Dissolved $O_3$) in plasma activated water prepared using multiple plasma device batch setup (refer figure 1 (b)). The statistically significant difference ($p < 0.05$) among the different properties of PAW groups with time (mean ± standard deviation) $\mu \pm \sigma$ is shown by different lowercase letters.

3.4 Properties of PAW prepared in continuous process

Figure 6 showed the properties of PAW when prepared using continuous processes. Similar to the batch process, the properties of PAW showed a similar trend with increasing plasma-water exposure time. However, one exception was observed in the form of dissolved $O_3$ concentration. In which, the dissolved $O_3$ concentration increases with increasing plasma-water exposure time. This was due to the low reactivity (lower oxidizing tendency and higher pH) of PAW in continuous processes which limits the reaction of dissolved $O_3$ with other RONS in PAW. Therefore, the decrease in dissolved $O_3$ concentration was not observed.

The chosen plasma-water exposure time was 10 hours for a continuous process which was significantly higher compared to the batch process (4 hours). Since, the volume of water for the continuous process was 20 liters that was larger compared to the batch process (2 liters). As the multiple plasma device was used for both the batch and continuous process. Therefore, a higher plasma-water exposure time was used in continuous processes compared to batch processes.

For a plasma-water exposure time of 10 hours, the percentage decrease in the pH of PAW was 53.7% compared to control (water with no plasma treatment, t = 0 hours). Simultaneously, the increase in ORP of PAW was 108.7%. Moreover, the TDS and EC of PAW after a plasma-water exposure time of 10 hours reached the value of 340 ppm and 850 µS cm$^{-1}$, respectively. The measured maximum concentration of $NO_3^-$ ions, $NO_2^-$ ions, $H_2O_2$, and dissolved $O_3$ after 10 hours of plasma-water exposure were given as 93.5 mg L$^{-1}$, 7.5 mg L$^{-1}$, 4.7 mg L$^{-1}$, and 4.7 mg L$^{-1}$, respectively.

As the reactivity (oxidizing tendency) and dissolved RONS concentration in PAW prepared in a continuous process were lower than the batch process. Hence, this low reactive PAW has applications in seed germination and plant growth and as a nitrogen source in agriculture applications, etc.(17, 18, 21, 22, 53) Since, as discussed, the high reactive PAW has applications in microbial inactivation. Hence, seeds and plants' exposure to high reactive PAW may damage the seeds and plants themselves (53). Therefore, control over the PAW properties becomes extremely important while preparing PAW for different applications.

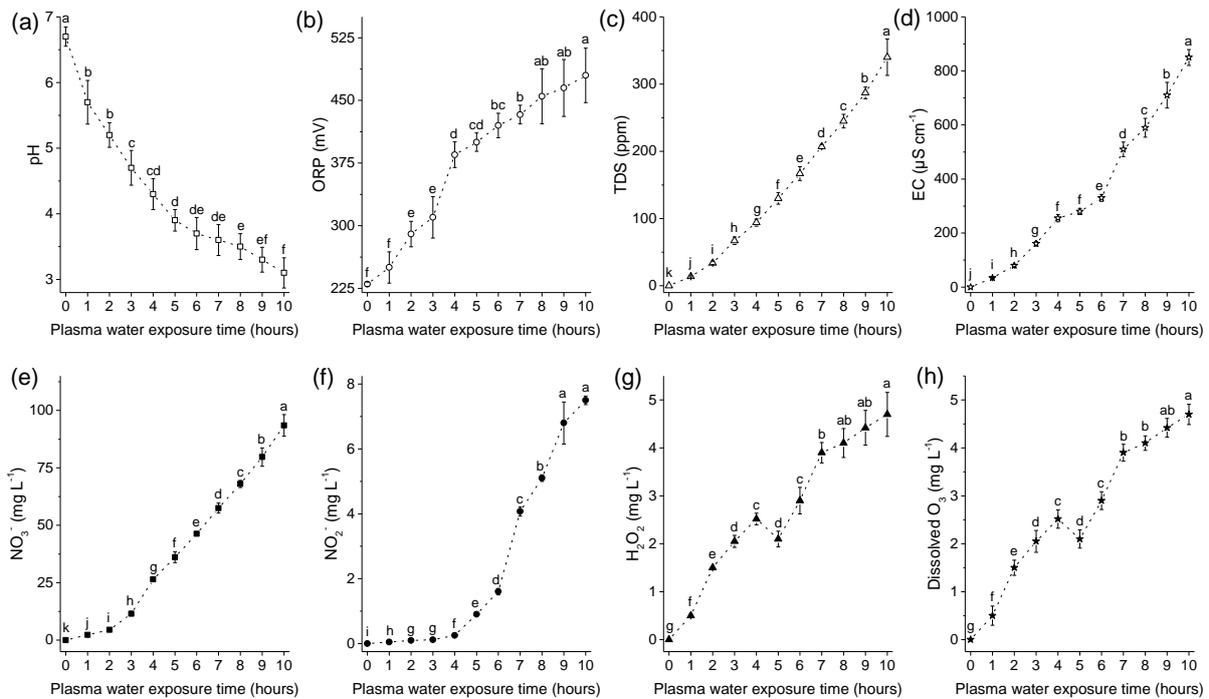

Figure 6. Physicochemical properties ((a) pH, (b) oxidation-reduction potential (ORP), (c) total dissolved solids (TDS), and (d) Electrical conductivity (EC)) and reactive-oxygen nitrogen species concentration ((e) $NO_3^-$ ions, (f) $NO_2^-$ ions, (g) $H_2O_2$, and (h) Dissolved $O_3$) in plasma activated water prepared using multiple plasma device continuous setup (refer figure 1 (c)). The statistically significant difference ($p < 0.05$) among the different properties of PAW groups with time (mean ± standard deviation) $\mu \pm \sigma$ is shown by different lowercase letters.

## 4. Conclusion

The present work discussed a multiple plasma device setup to produce plasma activated water. This multiple plasma device produces PAW in batch and continuous manner. Both the batch and continuous PAW setup are designed in such a way as to optimize the dissolution of gases reactive species in water. Moreover, the reactive species carried by effluent gases after plasma-water exposure are trapped in water in both the batch and continuous PAW setup. Hence, it reduced the environmental pollutants such as $NO_x$ and gases $O_3$, etc. which comes out effluent gases after plasma-water exposure. In the batch process, a high reactive PAW is produced with a total volume of 2 liters. In the continuous process, a relatively lower reactive PAW is produced with a total volume of 20 liters. The high reactive PAW has enhanced physicochemical properties and RONS concentration. Hence, it has applications in pathogen inactivation, food preservation, cancer cell inactivation, and surface disinfection, etc. Moreover, less reactive PAW has moderate physicochemical properties and lower RONS concentration compared to a high reactive PAW. Therefore, it has applications in seeds germination, plant growth, and as a nitrogen source for numerous agriculture and aquaculture applications, etc.

## Acknowledgments

This work was supported by the Department of Atomic Energy (Government of India) graduate fellowship scheme (DGFS). The authors sincerely thank O. R. Kaila, and Mr. Nimish for providing constant support and useful suggestions during this work.

## Data availability statement

The data that support the findings of this study are available upon reasonable request from the authors.

## Conflict of interests

The authors declare that there are no conflicts of interests.

## Authors' contributions

All authors contributed to the study conception and design. Material preparation, data collection, and analysis were performed by Vikas Rathore. Design and development of power supply was done by Chirayu Patil. The first draft of the manuscript was written by Vikas Rathore, and all authors commented on previous versions of the manuscript. All authors read and approved the final manuscript.

**ORCID iDs**

Vikas Rathore https://orcid.org/0000-0001-6480-5009

**Annexure**

Table A1: Reactions of formation of various reactive oxygen-nitrogen species in PAW through plasma-water interaction.

| Reaction | Rate constant (k) | Eq. no. |
|---|---|---|
| $N_2\ (g) + e^- \rightarrow 2N\ (g) + e^-$ | $6.3 \times 10^{-6}\ T_e^{-1.6}\ e^{-9.8/T_e}$ cm$^3$ s$^{-1}$ | 2 |
| $H_2O\ (g) + e^- \rightarrow OH\ (g) + H\ (g) + e^-$ | $2.6 \times 10^{-12}$ cm$^3$ s$^{-1}$ | 3 |
| $H_2O\ (g) + N_2(A)\ (g) \rightarrow OH\ (g) + H\ (g) + N_2\ (g)$ | $4.2 \times 10^{-11}$ cm$^3$ s$^{-1}$ | 4 |
| $N\ (g) + O_2\ (g) \rightarrow NO\ (g) + O\ (g)$ | $8.5 \times 10^{-17}$ cm$^3$ s$^{-1}$ | 5 |
| $O(g) + NO(g) + M(g) \rightarrow NO_2(g) + M(g)$ | $9.0 \times 10^{-32}$ cm$^6$ s$^{-1}$ | 6 |
| $O\ (g) + O_2\ (g) \rightarrow O_3\ (g)$ | $1.7 \times 10^{-12}$ cm$^3$ s$^{-1}$ | 7 |
| $OH\ (g) + OH\ (g) \rightarrow H_2O_2\ (g)$ | $2.6 \times 10^{-11}$ cm$^3$ s$^{-1}$ | 8 |
| $OH\ (g) + O_3\ (g) \rightarrow HO_2\ (g) + O_2\ (g)$ | $1.9 \times 10^{-12}$ cm$^3$ s$^{-1}$ | 9 |
| $OH\ (aq.) + NO\ (aq.) \rightarrow \mathbf{NO_2^-}\ (aq.) + H^+(aq.)$ | $1.0 \times 10^{10}$ M$^{-1}$ s$^{-1}$ | 10 |
| $\mathbf{O_3}\ (aq.) + \mathbf{NO_2^-}\ (aq.) \rightarrow O_2\ (aq.) + \mathbf{NO_3^-}\ (aq.)$ | $2.5 \times 10^5$ M$^{-1}$ s$^{-1}$ | 11 |
| $OH\ (aq.) + OH\ (aq.) \rightarrow \mathbf{H_2O_2}\ (aq.)$ | $5.0 \times 10^9$ M$^{-1}$ s$^{-1}$ | 12 |
| $OH\ (aq.) + \mathbf{O_3}\ (aq.) \rightarrow O_2\ (aq.) + HO_2\ (aq.)$ | $1.0 \times 10^8$ M$^{-1}$ s$^{-1}$ | 13 |
| $HO_2\ (aq.) + HO_2\ (aq.) \rightarrow O_2\ (aq.) + \mathbf{H_2O_2}\ (aq.)$ | $1.0 \times 10^6$ M$^{-1}$ s$^{-1}$ | 14 |
| $NO_2\ (aq.) + NO\ (aq.) + H_2O\ (aq.) \rightarrow \mathbf{2NO_2^-}\ (aq.) + 2H^+\ (aq.)$ | $2.0 \times 10^8$ M$^{-1}$ s$^{-1}$ | 15 |
| $2NO_2\ (aq.) + H_2O\ (aq.) \rightarrow \mathbf{NO_3^-}\ (aq.) + \mathbf{NO_2^-}\ (aq.) + 2H^+\ (aq.)$ | $0.5 \times 10^8$ M$^{-1}$ s$^{-1}$ | 16 |
| $\mathbf{H_2O_2}\ (aq.) + OH\ (aq.) \rightarrow H_2O\ (aq.) + O_2^-\ (aq.) + H^+\ (aq.)$ | $2.7 \times 10^7$ M$^{-1}$ s$^{-1}$ | 17 |
| $H_2O\ (aq.) + HO_2\ (aq.) + O_2^-\ (aq.)$ $\rightarrow O_2\ (aq.) + \mathbf{H_2O_2}\ (aq.) + HO^-\ (aq.)$ | $9.7 \times 10^7$ M$^{-1}$ s$^{-1}$ | 18 |
| $\mathbf{NO_2^-}\ (aq.) + H_2O_2\ (aq.) + H^+\ (aq.) \rightarrow ONOOH\ (aq.) + H_2O\ (aq.)$ | $1.1 \times 10^3$ M$^{-1}$ s$^{-1}$ | 19 |
| $NO_2^- \ (aq.) + OH \ (aq.) + H^+ \ (aq.) \rightarrow NO_3^- \ (aq.) + 2H^+ \ (aq.)$ | $5.3 \times 10^9$ M$^{-1}$ s$^{-1}$ | 20 |
| $ONOOH \ (aq.) \rightarrow NO_3^- \ (aq.) + H^+ \ (aq.)$ | 0.9 s$^{-1}$ | 21 |